\documentclass[letterpaper,10pt]{article}
\usepackage{osameet2}

\usepackage{tabularx}
\usepackage{amsmath,amssymb}
\usepackage{epstopdf}
\usepackage{graphicx} 
\usepackage{floatrow}
\usepackage{color}
\usepackage{svg}
\usepackage[acronym]{glossaries}
\usepackage{url}
\usepackage{hyperref}
\usepackage{caption}
\usepackage{adjustbox}
\usepackage{comment}
\usepackage{wrapfig}
\usepackage{fancyhdr}

\usepackage{tikz,pgfplots,subfig, filecontents}

\pgfplotsset{compat=1.15}

\colorlet{figyellow}{yellow!40!white}
\colorlet{figred}{red!40!white}
\colorlet{figblue}{blue!20!white}
\colorlet{figgreen}{green!30!white}
\colorlet{figgray}{black!10!white}

\definecolor{myGreen}{RGB}{60,179,113}
\definecolor{myRed}{RGB}{255,48,48}
\definecolor{myPurple}{RGB}{128,0,128}
\definecolor{myBlue}{RGB}{72,118,255}
\definecolor{myOrange}{RGB}{138,54,15}
\definecolor{InfinBlue}{RGB}{42,42,61}
\definecolor{Senna}{RGB}{255,0,0}
\definecolor{myGrey}{RGB}{224,224,224}
\definecolor{myWhite}{RGB}{255,255,255}

\newfloatcommand{capbtabbox}{table}[][\FBwidth]

\makeglossaries
\newacronym{dal}{DAL}{Dallas}
\newacronym{mem}{MEM}{Memphis}
\newacronym{cha}{CHA}{Chattanooga}
\newacronym{zrhh}{400ZR w/Regen}{400ZR Hop-by-Hop P2P}
\newacronym{zreh}{400ZR+ w/Express}{400ZR+ P2P}
\newacronym{xr}{400G XR}{400G P2MP}
\newacronym{adc}{ADC}{Analog-to-Digital Converterss}
\newacronym{adsl}{ADSL}{Asymmetric Digital Subscriber Line}
\newacronym{am}{AM}{Aggregation Managers}
\newacronym{api}{API}{Application Programming Interface}
\newacronym{ar}{AR}{Augmented Reality}
\newacronym{ase}{ASE}{Amplified Spontaneous Emission}
\newacronym{asic}{ASIC}{Application-specific Integrated Circuits}
\newacronym{b2b}{B2B}{Back-to-back}
\newacronym{b5g}{B5G}{Beyond 5G}
\newacronym{bt}{BT}{British Telecom}
\newacronym{bw}{BW}{Bandwidth}
\newacronym{capex}{CAPEX}{Capital Expenditure}
\newacronym{cd}{CD}{Chromatic Dispersion}
\newacronym{cdc}{CDC}{Colorless-Directionless-Contentionless}
\newacronym{cfo}{CFO}{Carrier Frequency Offset}
\newacronym{cfp2}{CFP2}{Centum Form-Factor Pluggable 2}
\newacronym{cut}{CUT}{Channel Under Test}
\newacronym{cmis}{CMIS}{Common Management Interface Specification}
\newacronym{cmos}{CMOS}{Complementary Metal–oxide–semiconductor}
\newacronym{co}{CO}{Central Office}
\newacronym{cpb}{CpB}{Cost per Transported Bit}
\newacronym{cpe}{CPE}{Carrier Phase Estimation}
\newacronym{cpu}{CPU}{Central Processing Unit}
\newacronym{cr}{CR}{Carrier Recovery}
\newacronym{cran}{CRAN}{Centralized Radio Access Network}
\newacronym{csp}{CSP}{Content Service Provider}
\newacronym{cwdm}{CWDM}{Coarse Wavelength Division Multiplexing}
\newacronym{cu}{CU}{Centralized Unit}
\newacronym{daa}{DAA}{Distributed Access Architecture}
\newacronym{dac}{DAC}{Digital-to-Analog Converters}
\newacronym{dacc}{$D_{\text{acc}}$}{Accumulated Dispersion}
\newacronym{dc}{DC}{Data Center}
\newacronym{dci}{DCI}{Data Center Interconnect}
\newacronym{dd}{DD}{Direct Detection}
\newacronym{dm}{DM}{Domain Manager}
\newacronym{dmt}{DMT}{Discrete Multitone}
\newacronym{dl}{DL}{Downlink}
\newacronym{dpu}{DPU}{Data Processing Unit}
\newacronym{dscm}{DSCM}{Digital Subcarrier Multiplexing}
\newacronym{dsc}{DSC}{Digital Subcarrier}
\newacronym{dsp}{DSP}{Digital Signal Processing}
\newacronym{du}{DU}{Distributed Unit}
\newacronym{dwdm}{DWDM}{Dense Wavelength Division Multiplexing}
\newacronym{edfa}{EDFA}{Erbium Doped Fiber Amplifier}
\newacronym{edf}{EDF}{Erbium Doped Fiber}
\newacronym{eepn}{EEPN}{Enhanced Equalization Phase Noise}
\newacronym{e2e}{E2E}{End-to-end}
\newacronym{eo}{E/O}{Electro-optical}
\newacronym{eon}{EON}{Elastic Optical Network}
\newacronym{evb}{EVB}{Evaluation Board}
\newacronym{fdm}{FDM}{Frequency Division Multiplexing}
\newacronym{fec}{FEC}{Forward Error Correction}
\newacronym{ff}{FF}{Form Factor}
\newacronym{fir}{FIR}{Finite Impulse Response}
\newacronym{foadm}{FOADM}{Fixed Optical Add/Drop Multiplexers}
\newacronym{ftth}{FTTH}{Fiber-to-the-Home}
\newacronym{ftta}{FTTA}{Fiber-to-the-Antenna}
\newacronym{fw}{FW}{Firmware}
\newacronym{gcc}{GCC}{General Communication Channel}
\newacronym{gpon}{GPON}{Gigabit Ethernet Passive Optical Network}
\newacronym{gprs}{GPRS}{General Packet Radio Service}
\newacronym{gui}{GUI}{Graphical User Interface}
\newacronym{has}{H\&S}{Hub and Spoke}
\newacronym{hd}{HD}{Hard Decision}
\newacronym{hl}{HL}{Hierarchy Layer}
\newacronym{ia}{IA}{Implementation agreement}
\newacronym{icp}{ICP}{Internet Content Provider}
\newacronym{imdd}{IM-DD}{Intensity Modulation Direct-Detection}
\newacronym{inp}{InP}{Indium Phosphide}
\newacronym{ip}{IP}{Internet Protocol}
\newacronym{ipodwdm}{IPoDWDM}{IP over DWDM}
\newacronym{ipm}{IPM}{Intelligent Pluggable Manager}
\newacronym{iot}{IoT}{Internet of Things}
\newacronym{ipt}{IPT}{Intelligent Pluggable Transceiver}
\newacronym{iqmzm}{IQ-MZM}{IQ Mach-Zehnder Modulator}
\newacronym{itrs}{ITRS}{International Technology Roadmap for Semiconductors}
\newacronym{itu}{ITU}{International Telecommunication Union}
\newacronym{lan}{LAN}{Local Area Network}
\newacronym{leaf}{LEAF}{Large Effective Area Fiber}
\newacronym{lldp}{LLDP}{Link Layer Discovery Protocol}
\newacronym{lo}{LO}{Local Oscillator}
\newacronym{lti}{LTI}{Linear Time Invariant}
\newacronym{lw}{LW}{Linewidth}
\newacronym{man}{MAN}{Metro Area Network}
\newacronym{mcu}{MCU}{Microcontroller unit}
\newacronym{mdio}{MDIO}{Management Data Input/Output}
\newacronym{mec}{MEC}{Multi-Access Edge Computing}
\newacronym{m2m}{M2M}{Machine-to-machine}
\newacronym{mimo}{MIMO}{Multiple Input Multiple Output}
\newacronym{mis}{MIS}{Management Interface Specification}
\newacronym{mpls}{MPLS}{Multi-Protocol Label Switching}
\newacronym{msa}{MSA}{Multi-Source Agreement}
\newacronym{mzm}{MZM}{Mach-Zehnder Modulator}
\newacronym{mzmd}{MZMD}{Mach-Zehnder Modulator Driver}
\newacronym{ncg}{NCG}{Net Coding Gain}
\newacronym{nid}{NID}{Network Interface Device}
\newacronym{nic}{NIC}{Network Interface Card}
\newacronym{nm}{NM}{Network Management}
\newacronym{odn}{ODN}{Optical Distribution Network}
\newacronym{oeo}{OEO}{Opto-Electro-Optical}
\newacronym{oe}{O/E}{Opto-electrical}
\newacronym{ofdm}{OFDM}{Orthogonal Frequency-division Multiplexing}
\newacronym{ogw}{OGW}{Optical Gateway}
\newacronym{oif}{OIF}{Optical Internetworking Forum}
\newacronym{olt}{OLT}{Optical Line Termination}
\newacronym{ols}{OLS}{Optical Line System}
\newacronym{ont}{ONT}{Optical Network Terminal}
\newacronym{ook}{OOK}{On-off Keying}
\newacronym{oon}{OON}{Open Optical Networks}
\newacronym{opex}{OPEX}{Operational Expenditure}
\newacronym{osnr}{OSNR}{Optical Signal-to-Noise-Ratio}
\newacronym{os}{OS}{Operating System}
\newacronym{osa}{OSA}{Optical Spectrum Analyzer}
\newacronym{osc}{OSC}{Optical Service Channel}
\newacronym{osfp}{OSFP}{Octal Small Form Factor Pluggable}
\newacronym{osi}{OSI}{Open System Interconnection}
\newacronym{otn}{OTN}{Optical Transport Network}
\newacronym{pam4}{4-PAM}{4 - Pulse Amplitude Modulation}
\newacronym{psd}{PSD}{Power Spectral Density}
\newacronym{pdh}{PDH}{Plesiochronous Digital Hierarchy}
\newacronym{pmd}{PMD}{Polarization Mode Dispersion}
\newacronym{pmo}{PMO}{Present Mode of Operation}
\newacronym{p2mp}{P2MP}{Point-to-Multipoint}
\newacronym{p2p}{P2P}{Point-to-Point}
\newacronym{pcs}{PCS}{Probabilistic Constellation Shaping}
\newacronym{pm}{PM}{Polarization Multiplexing}
\newacronym{pic}{PIC}{Photonic Integrated Circuit}
\newacronym{pon}{PON}{Passive Optical Network}
\newacronym{qam}{QAM}{Quadrature Amplitude Modulation}
\newacronym{16qam}{16QAM}{16 Quadrature Amplitude Modulation}
\newacronym{8qam}{8QAM}{8-Quadrature Amplitude Modulation}
\newacronym{4qam}{4QAM}{4-Quadrature Amplitude Modulation}
\newacronym{qpsk}{QPSK}{Quadrature Phase Shift Keying}
\newacronym{qsfp}{QSFP}{Quad Small Form Factor Pluggable}
\newacronym{qsfpdd}{QSFP-DD}{Quad Small Form Factor Pluggable Double Density}
\newacronym{ddfp}{DD}{Double Density}
\newacronym{ran}{RAN}{Radio Access Network}
\newacronym{roadm}{ROADM}{Reconfigurable Optical Add and Drop Multiplexer}
\newacronym{roi}{ROI}{Return on Investment}
\newacronym{ron}{RON}{Routed Optical Networks}
\newacronym{rpd}{RPD}{Remote PHY Device}
\newacronym{rx}{Rx}{Receiver}
\newacronym{ru}{RU}{Radio Unit}
\newacronym{sbvt}{S-BVT}{Sliceable Bandwidth Variable Transponder}
\newacronym{sc}{SC}{Subcarrier}
\newacronym{sc-on}{SCON}{Software Configurable Optical Networks}
\newacronym{sdh}{SDH}{Synchronous Digital Hierarchy}
\newacronym{sdn}{SDN}{Software Defined Networking}
\newacronym{sd}{SD}{Soft Decision}
\newacronym{se}{SE}{Spectral Efficiency}
\newacronym{sige}{SiGe}{Silicon Germanium}
\newacronym{siph}{SiPh}{Silicon photonics}
\newacronym{snic}{SmartNIC}{Smart Network Interface Card}
\newacronym{snr}{SNR}{Signal-to-Noise-Ratio}
\newacronym{sonet}{SONET}{Synchronous Optical Networking}
\newacronym{sota}{SotA}{State-of-the-art}
\newacronym{soa}{SOA}{Semiconductor Optical Amplifier}
\newacronym{spi}{SPI}{Serial Peripheral Interface}
\newacronym{sr}{$R_s$}{Symbol Rate}
\newacronym{ssmf}{SMF-28}{Single Mode Fiber}
\newacronym{tca}{TCA}{Threshold Crossing Alarm}
\newacronym{tco}{TCO}{Total Cost of Ownership}
\newacronym{tdm}{TDM}{Time Division Multiplexing}
\newacronym{tdma}{TDMA}{Time Division Multiple Access}
\newacronym{tec}{TEC}{Thermoelectric Cooler}
\newacronym{tia}{TIA}{Trans-Impedance Amplifier}
\newacronym{tim}{TIM}{Telecom Italia Mobile}
\newacronym{tosnr}{TOSNR}{Transmitted OSNR}
\newacronym{trosa}{TROSA}{Transmitter-Receiver Optical Sub-Assembly} 
\newacronym{tx}{Tx}{Transmitter}
\newacronym{ul}{UL}{Uplink}
\newacronym{ull}{ULL}{Ultra-Low-Loss Fiber}
\newacronym{upf}{UPF}{User Plane Function}
\newacronym{vco}{VCO}{Voltage-Controlled Oscillator}
\newacronym{vdu}{VDU}{Virtualized Distributed Unit}
\newacronym{vlan}{VLAN}{Virtual LAN}
\newacronym{vrar}{VR/AR}{Virtual and Augmented Reality}
\newacronym{wan}{WAN}{Wide Area Network}
\newacronym{wdm}{WDM}{Wavelength Division Multiplexing}
\newacronym{wss}{WSS}{Wavelength Selective Switch}
\begin{document}
\fancyhf{}
\renewcommand{\headrulewidth}{0pt}
\fancyfoot[c]{}
\fancypagestyle{FirstPage}{
\lfoot{
979-8-3503-7732-3/24/\$31.00 \copyright2024 IEEE
}
}

\title{On the impact of VR/AR applications on optical transport networks: First experiments with Meta Quest 3 gaming and conferencing applications}

\author{
C. de Quinto$^1$, 
A. Navarro$^1$,
G. Otero$^1$,
N. Koneva$^1$,
J. A. Hern\'{a}ndez$^1$,\\
M. Quagliotti$^3$
A. S\'{a}nchez-Macian$^1$,
F. Arpanaei$^1$,
P. Reviriego$^2$,
\'{O}. Gonz\'{a}lez de Dios$^4$,
J. M. Rivas-Moscoso$^4$,
E. Riccardi$^3$,
D. Larrabeiti$^1$,
}
\address{$^1$Univ. Carlos III de Madrid, Spain $^2$ Univ. Politecnica de Madrid, Spain $^3$ TIM, Italy, $^4$ Telefonica I+D, Spain}

\begin{abstract}
With the advent of next-generation AR/VR headsets, many of them with affordable prices, telecom operators have forecasted an explosive growth of traffic in their networks. Penetration of AR/VR services and applications is estimated to grow exponentially in the next few years. This work attempts to shed light on the bandwidth capacity requirements and latency of popular AR/VR applications with four different real experimental settings on the Meta Quest 3 headsets, and their potential impact on the network. 
\\
\\
\textbf{Keywords:} AR/VR; Metaverse; Traffic profiles; Meta Quest 3    .
\end{abstract}

\section{Introduction}
\thispagestyle{FirstPage}

In its latest report~\cite{ieee_fourkey}, the IEEE Standards Association has uncovered four key future-focused trends expected to shape the foundational technology landscape for 2024 and beyond: Evolution of the Metaverse, Building Trust with Data Governance, Child Safety Online, and Advances in Quantum Computing and New Applications. 

Indeed, the global AR/VR and VR headset market size reached US\$ 16.6 Billion in 2023 and is expected to grow at a compound annual growth rate (CAGR) of 12.44\% during the period 2024-2032. Concerning market share, the number of AR/VR headsets is rapidly growing at an exponential pace with Meta Quest 3 and Playstation VR Headset being the sales leaders. Recently Apple Vision Pro has appeared in the market selling all their stock. Other brands like Pico and HTC have a reasonable market share, while other consumer brands like Xiaomi and Samsung have recently announced plans for dealing AR/VR products sometime in 2024. These devices have penetrated 1-2\% of households in North America, the EU, and China.

The Metaverse opens new possibilities with a wide range of applications and use cases: (1) 360 gaming (either online or local), (2) entertainment (concerts, sports events), (3) training and education, (4) remote healthcare, (5) immersive tourism, etc. Some of these use cases and related network-based experiments can be found in~\cite{metaverse_hype,metaverse_intro}. At a high level, the metaverse can be categorized into these three business sectors: Industrial, Enterprise, and Consumer.

A good example of Industrial Metaverse, as noted in~\cite{ETS23},  can be Tactile Internet for Remote Surgery: This use case delves into the challenges and requirements for remote telesurgery, including ultra-low latency (below 1 ms), high reliability (up to 99.9999999\%) for UHD medical video over non-public networks (NPNs)~\cite{ETS22}, and massive data rates (up to 1 Tbit/s) to support applications like AR and Holographic Type Communication (HTC)~\cite{ITU20-1}. Regarding Enterprise Metaverse, Academic/Professional e-Learning explores the potential of immersive technologies VR/AR in enhancing educational experiences, both in academic and professional settings. It discusses the requirements for high-quality video streaming (up to 2.35 Gbps for eXtended Reality XR), ultra-low network latency (haptic response time to 5.5 ms), and scalability to support a large number of students simultaneously~\cite{IOW21-2}. Finally, as an example of use cases with high development potential within the Consumer Metaverse, Virtual Tourism in Smart City (Consumer Metaverse) focuses on the integration of virtual and augmented reality experiences in the context of smart cities, enabling immersive tourism experiences and metaverse-driven services for citizens and visitors. These services require efficient mobility and security solutions, with strict requirements for low latency (20 ms for Ultra-Reliable Low Latency Communications (URLLC)), high data mobility (10 km/h for AR/VR, 0,5 km/h for Telepresence), data rate (from 40 to 600 Mbit/s)  and reliability (up to 99.9999\%) to support these applications~\cite{5GEVE}.


While new market opportunities appear with AR/VR products, integrating these technologies necessitates significant network upgrades to handle the increased bandwidth and latency demands they impose. AR/VR applications heavily rely on streaming high-resolution visuals and spatial data, requiring significantly more bandwidth than traditional voice or video calls. 5G networks with near-Gbps speeds are crucial for seamless AR/VR experiences. Regarding latency, high delay and delay variation (aka jitter) in data transmission can cause nausea and disorientation in VR environments. Real-time interaction requires ultra-low latency networks, pushing the boundaries of current telecommunication infrastructure. Thus, the proliferation of AR/VR devices will significantly increase overall network traffic, demanding higher capacity to avoid congestion and maintain consistent performance. 

Hence, telecom operators need to prepare for a possible exponential growth in the traffic injected by these applications into their networks and the related network requirements demanded by them. Among other aspects, telcos will be required to update their infrastructure, investing heavily in fiber optic rollouts, edge computing infrastructure, and 5G-A/6G deployments.


\section{State of the Art}

Recent studies have focused into various aspects of AR/VR applications, highlighting critical factors such as network performance, privacy concerns, interaction with sensor networks, security vulnerabilities, and cloud-rendering architectures. In~\cite{casasnovas2024experimental}, the authors explore the relationship between Wi-Fi performance and the quality of VR streaming experiences. The study investigates the characteristics of VR traffic over Wi-Fi networks, particularly focusing on sustained network performance and the impact of frame rate on Wi-Fi efficiency. Notably, it identifies a segmentation mechanism in WebRTC-based services that affects Wi-Fi airtime consumption.

The study carried out in~\cite{jarin2023behavr} addresses privacy risks associated with VR platforms, emphasizing the collection of sensitive data by VR sensors and the potential for user identification. The study introduces BEHAVR, a framework for analyzing sensor data across VR applications, demonstrating high accuracy in user identification. This highlights the importance of considering privacy implications in AR/VR research. 

In~\cite{makolkina:hal-01675414}, Makolkina et al. explore the interaction between augmented reality and flying ubiquitous sensor networks (FUSN), emphasizing the need for new traffic patterns to ensure the quality of experience. The study proposes a novel traffic pattern capturing service space, environment, and user behavior models, suggesting advancements in AR technologies.

The authors of \cite{Leeking} focus on security and privacy issues in mobile AR applications, specifically regarding the inference of user location based on network traffic patterns. The study demonstrates a side-channel attack against a popular AR application, highlighting vulnerabilities in location-based AR services and advocating for mitigation strategies. 

In~\cite{Schulzmodeling}, the authors investigate cloud-rendering architectures for AR on lightweight glasses, emphasizing the challenges of low latency and high data rates in wireless networks. The study proposes a realistic traffic model based on video data analysis, aiming to assess network performance in cloud-rendered AR scenarios.

In contrast to the aforementioned studies, our paper provides a comprehensive overview of traffic patterns observed in various AR/VR applications, including onsite video gaming, external rendering video gaming, and virtual reality video streaming. While~\cite{casasnovas2024experimental} and~\cite{Schulzmodeling} address network performance in VR streaming and cloud-rendering architectures respectively, our study extends this by examining traffic profiles across different AR/VR scenarios. Furthermore, we complement~\cite{jarin2023behavr} by focusing on traffic characteristics rather than privacy concerns, thereby contributing to a holistic understanding of AR/VR application dynamics. Moreover, while~\cite{makolkina:hal-01675414} explores new traffic patterns for augmented reality, our study extends this by examining traffic profiles in diverse AR/VR contexts.

In this document, we present multiple setups for studying the impact of different AR/VR applications on the network. In particular, we have collected packet traces in several application scenarios and measured latency and bandwidth requirements as they traverse the access networks. We observe that different applications and setups show various traffic profiles, in most cases similar to high-resolution video-streaming (like 4K or 8K resolution video). 

We believe that our research provides valuable insights into the traffic patterns critical for ensuring the quality of AR/VR experiences, thereby advancing the understanding of network requirements and performance optimization in immersive applications. The remainder of this document is organized as follows: Section~\ref{sec:scenarios} overviews the different measurement setups for the experiments, whose results are explained in Section~\ref{sec:experiments}. Finally, Section~\ref{sec:conclusions} concludes this article with a summary of its main findings and conclusions.

\section{AR/VR setups and experiments}
\label{sec:scenarios}

\begin{wrapfigure}{r}{0.6\textwidth}
  \begin{center}
    \includegraphics[width=0.99\textwidth]{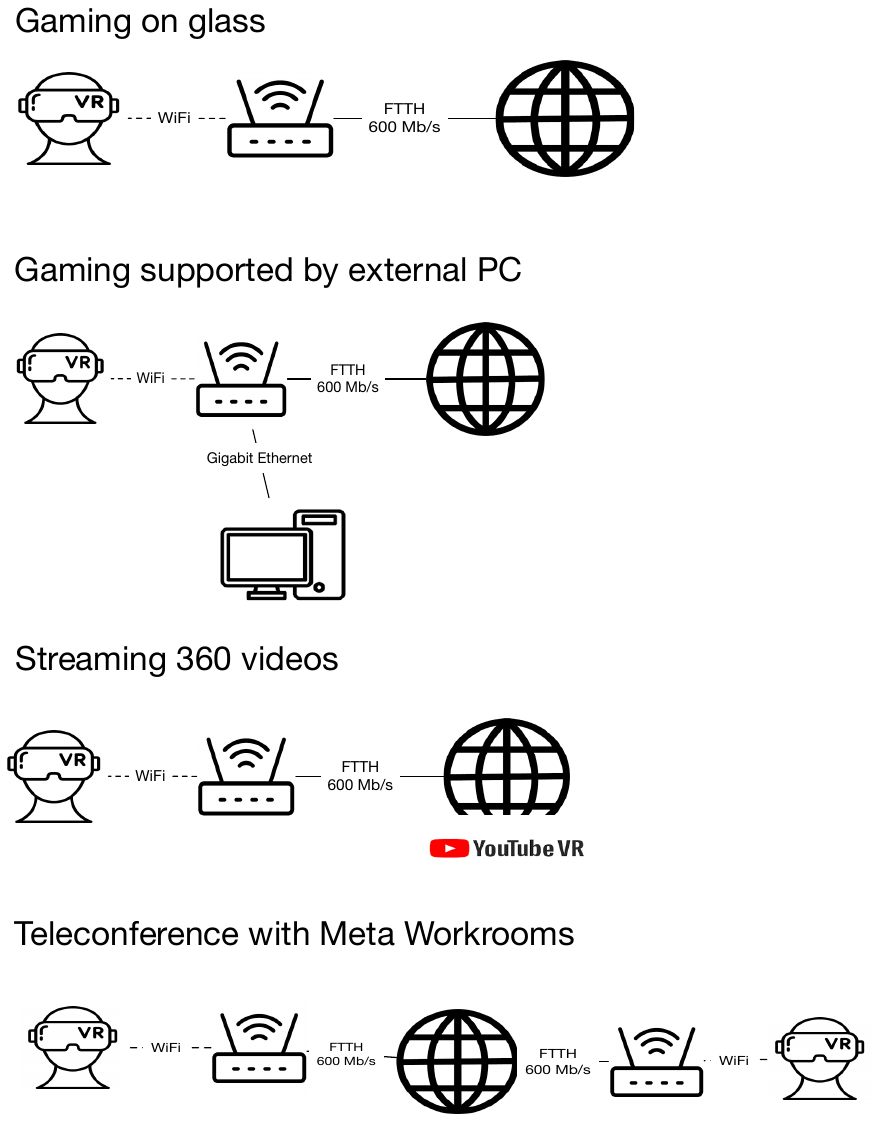}
  \end{center}
 \caption{Overview of all deployed scenarios}
	\label{fig:scenariosaall}
\end{wrapfigure}

\textbf{\#1 Gaming on the headset (Hyper Dash game):} This scenario represents the use case where a consumer plays a low-resolution game that runs directly on the MetaQuest3 hardware. (See Fig.~\ref{fig:scenariosaall}); the computational power required for rendering this game is affordable and the headset itself can run the game without external support. Captured traces reveal that Kbps traffic (about 50 packet/s with periodic spikes) goes from the headset to the Internet, mainly the movements of the controllers and the keystrokes and commands (shoots, position, etc), that is, important information for online gaming with other players around the world. This behavior has been previously observed in previous online gaming studies (without AR/VR headset), see~\cite{manzano_2012}.

\textbf{\#2 Gaming supported by external computer:} Fig.~\ref{fig:scenariosaall} shows a second gaming scenario where the game does not run on the headset but on a physically closed desktop computer. Here, the game demands high computational power (GPUs for graphics rendering), thus an external computer is needed to properly operate. The traces reveal a continuous flow of around 157 Mb/s from the local PC station towards the AR/VR headset, and again a few Kb/s towards the Internet.

\textbf{\#3 Streaming 360 videos (YoutubeVR):} In this case, the user is watching a 360-video (Youtube VR application) streamed over the Internet. Fig.~\ref{fig:scenariosaall} details the scenario and bitrate arrival from the streaming server (Youtube VR), showing an average bitrate of 60 Mb/s downstream; this is approximately the typical bitrate of 3x a classical 2K (i.e. 2048 × 1080) video stream.

\textbf{\#4 Collaborative business meeting (with Meta Workrooms)} In the fourth scenario, two users from different cities in Madrid region (Spain) are having a teleconference using Meta Workrooms, an application for teleconference (see Fig.~\ref{fig:scenariosaall}). The two users see each other's avatar and exchange some files while talking and drawing diagrams with their fingers. In this case, the traffic exchanged between them is approximately 15 Mb/s.

\begin{figure}[htbp]
\centering
\includegraphics[width =\textwidth]{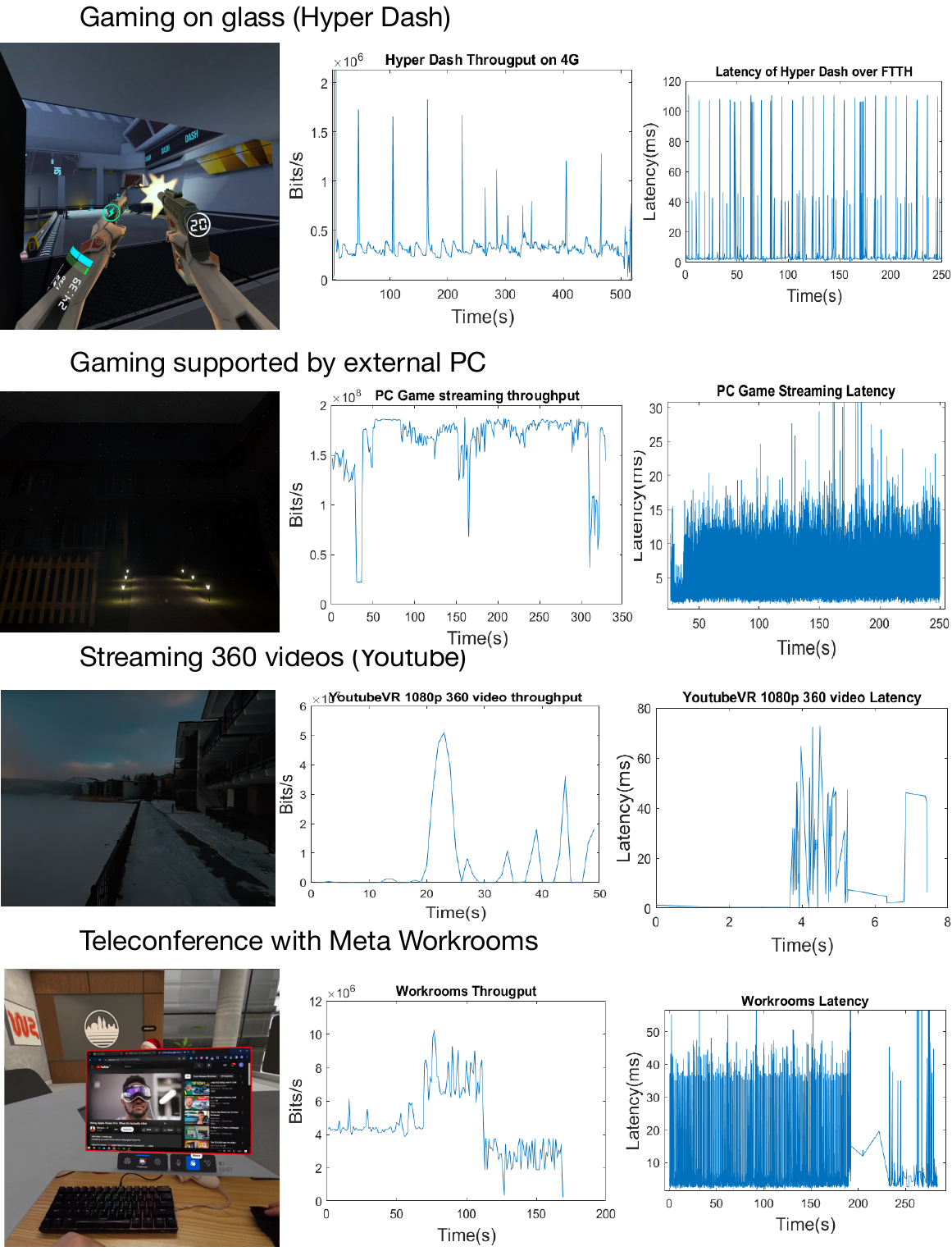}
\caption{Screenshot, bitrate and latency measured in all scenarios}
\label{fig:experiments}
\end{figure}

Fig.~\ref{fig:experiments} shows the bitrate (in bits/s) observed in the four scenarios

Table~\ref{tab:bitrates} presents a summary of the relevant bitrates, latency, and packet characteristics observed in each scenario. In the "gaming on headset" scenario, games are processed directly on the headset's hardware, resulting in lower game resolution and visual quality. This setup transmits only keystrokes and essential online gaming information to the Metro network. We noted that the game performs smoothly when the Internet connection is facilitated through a mobile phone (refer to the 4G row in the table), albeit with slightly higher latency (19.6 ms compared to 12.1 ms) compared to when the headset connects via WiFi to a home modem/router and then through FTTH (Fiber to the Home) to the Internet. In the second scenario, the game's processing is predominantly handled by an external PC at the user’s home, with the bitrates and latency figures reflecting the connection between the user and this external PC. The Internet traffic in this scenario closely mirrors that of the "gaming on headset" scenario. Regarding the 360-video streaming scenario to YoutubeVR, the bitrates vary based on the original video resolution and align with typical video streaming values. Lastly, in the Meta Workrooms case, the bitrate and latency metrics are in line with those expected for FullHD resolution video streaming to a server in the Metro segment. The session data from this Meta Workrooms connection has been archived and is accessible at \url{https://www.youtube.com/watch?v=cnQpoMByT3E}.

\begin{table}[!htbp]
    \centering
    \renewcommand{\arraystretch}{1.5} 
    \begin{tabularx}{\textwidth}{|*{7}{>{\centering\arraybackslash}X|}} 
    \hline
    \textbf{Scenario} &  \textbf{Avg Bitrate} (Mb/s) & \textbf{Sd Bitrate} (Mb/s) & \textbf{Avg Latency} (ms) & \textbf{Sd Latency} (ms) & \textbf{Avg Packet} (bit) & \textbf{Peak Packet} (bit) \\
    \hline
    Gaming on headset (4G) & 0.38 & 0.95 & 19.6 & 0.43 & 409 & 4200 \\ 
    Gaming on headset (FTTH) & 0.38 & 0.20 & 12.1 & 0.32 & 450 & 18000 \\
    Gaming on external PC & 157 & 32 & 6.2 & 0.01 & 4,556 & 64,054 \\
    360-video streaming & 60 & 12.8 & 23.6 & 1.7 & 1,147 & 2,800 \\
    Meta Workrooms & 15 & 1.95 & 12.5 & 0.14 & 700 & 18,000 \\
    \hline
    \end{tabularx}
    \caption{Summary of traffic profiles observed in all scenarios (Avg: average value, Sd: Standard deviation)}
    \label{tab:bitrates}
\end{table}




\section{Discussion: Impact on telecommunication infrastructure}
\label{sec:experiments}

The rise of metaverse applications presents significant challenges and opportunities for Telecom Operators, necessitating substantial investments to enhance network capabilities. These enhancements include increased bandwidth, reduced latency, and improved reliability, along with advanced Edge Cloud infrastructure to support these bandwidth-intensive applications.

Operators are currently focusing on several key AR/VR use cases, such as industrial remote operations, tele-education in remote or global settings with satellite network assistance, and entertainment applications like gaming and virtual tourism. These applications demand stringent network requirements to ensure a seamless and immersive user experience. Some of these requirements include~\cite{Holland2019}:
\begin{itemize}
    \item Extreme Low Latency: Essential for real-time social interactions and precise remote control in industrial settings, demanding network latencies ranging from 1 ms to less than 0.1 ms for the Radio Access Network (RAN). Specific latency needs vary by application, such as 0.5-2 ms for dynamic haptic feedback and up to 20 ms for Ultra-High Definition (UHD) video over significant distances.
    \item High Symmetrical Transmission Bandwidth: To support the complex data needs of AR/VR and HTC experiences, both downstream and upstream data rates will need to increase significantly. These rates range from tens or hundreds of Mbit/s for 4K video to 1 Tbit/s for holographic communications, ensuring a high-quality multisensory experience (QoE) for users.
    \item Service Availability: The reliability of these services is critical, with targets ranging from 99.999\% (five nines) to 99.9999999\% (nine nines) availability, indicating the network's expected operational excellence.
\end{itemize}


To meet the capacity requirements of these emerging use cases, more bandwidth needs to be made available in the RAN and other network segments, potentially through the exploitation of new frequency bands, massive MIMO processing, and innovative fiber technologies such as multi-core fibers or hollow-core fibers.

As AR/VR services and applications continue to gain traction, telecom operators must anticipate and address the associated challenges, including the need for higher bandwidth, lower latency, and improved reliability. This work aims to shed light on the bandwidth capacity requirements and latency of popular AR/VR applications through four different real experimental settings on the MetaQuest3 headsets, and its potential impact on the network.

\section{Summary and conclusions}
\label{sec:conclusions} 


Augmented Reality (AR) and Virtual Reality (VR) are novel technologies that have revolutionized the way we perceive and interact with the digital world. Both AR and VR alter our perception of reality but in different ways. AR overlays digital content onto the real world, enhancing our surroundings with additional information or virtual objects; VR on the contrary immerses users in a completely virtual environment, isolating them from the physical world and transporting them to simulated realms. Thanks to their impressive immersive features, these devices have penetrated 1-2\% households in North America, Europe, and China and are expected to grow exponentially in the next years. 

For these reasons, telecom operators must proactively prepare for the potentially explosive growth of these services and applications, equipping their networks to handle a scenario where Metaverse traffic becomes the leading source of streamed traffic.

This article has analyzed the traffic generated into the network by four applications (two games, 360 VR streaming, and Metaverse teleconference) in an AR/VR Meta Quest 3 scenario. 

We note that the network traffic closely mirrors that of online gaming or video streaming, potentially requiring additional bandwidth and stricter latency measures due to the complex immersive requirements of Metaverse applications.

\section*{Acknowledgements}

The following projects have contributed to the development of this work: Project ITACA (grant PDC2022-133888-I00) funded by Agencia Estatal de Investigación (MCIN, Spain); SNS project SEASON (grant 101096120) funded by EU Horizon-Europe Program. This work is also supported by the R\&D project PID2022-136684OB-C21 (Fun4Date) funded by the Spanish Ministry of Science and Innovation MCIN/AEI/ 10.13039/501100011033.





\bibliographystyle{unsrt}
\bibliography{references}

\end{document}